\documentclass[conference]{IEEEtran}
\IEEEoverridecommandlockouts
\usepackage{amsmath,amssymb,amsfonts}
\usepackage{algorithmic}
\usepackage{graphicx}
\usepackage{textcomp}
\usepackage{xcolor}
\usepackage{todonotes}
\usepackage{mwe}
\usepackage{subcaption}
\usepackage{listings}
\usepackage{xcolor}
\usepackage{url}
\presetkeys{todonotes}{inline}{}
\usepackage[backend=biber,style=ieee]{biblatex}
\def\charm{{Charm++}}
\def\charmpy{{Charm4Py}}
\def\mpi{{MPI}}
\def\mpipy{{mpi4py}}
\def\stampede{\textsc{stampede2}}
\def\summit{\textsc{summit}}

\def\BibTeX{{\rm B\kern-.05em{\sc i\kern-.025em b}\kern-.08em
    T\kern-.1667em\lower.7ex\hbox{E}\kern-.125emX}}
\begin{document}

\definecolor{belizehole}{HTML}{2980b9}
\definecolor{clouds}{HTML}{f4f8f9}
\definecolor{midnightblue}{HTML}{2c3e50}
\definecolor{pomegranate}{HTML}{c0392b}
\definecolor{royalblue}{HTML}{3867d6}
\definecolor{greensea}{HTML}{16a085}
\definecolor{nephritis}{HTML}{27ae60}
\definecolor{amethyst}{HTML}{9b59b6}

\lstset{
	basicstyle=\ttfamily,        
	breakatwhitespace=false,         
	breaklines=true,                 
	captionpos=b,                    
	commentstyle=\color{royalblue},    
	frame=single,                    
	keepspaces=true,                 
	keywordstyle=\color{amethyst},       
	numbers=none,                    
	numbersep=5pt,                   
	numberstyle=\tiny\color{gray}, 
	rulecolor=\color{black},         
	showspaces=false,                
	showstringspaces=false,          
	showtabs=false,                  
	stepnumber=2,                    
	stringstyle=\color{red},     
	tabsize=2,                       
	title=\lstname,                   
	belowskip=-10pt,
	numbers=left
}

\lstdefinestyle{interfaces}{
	float=t,
	floatplacement=t,
}

\title{Performance Evaluation of Python Parallel Programming Models: Charm4Py and mpi4py
}
\author{\IEEEauthorblockN{Zane Fink\IEEEauthorrefmark{1}, Simeng Liu\IEEEauthorrefmark{1}, Jaemin Choi\IEEEauthorrefmark{1}, Matthias Diener\IEEEauthorrefmark{1}, Laxmikant V.~Kale\IEEEauthorrefmark{1}\IEEEauthorrefmark{2}}
\IEEEauthorblockA{\IEEEauthorrefmark{1}Department of Computer Science, University of Illinois at Urbana-Champaign, Urbana, Illinois, USA}
\IEEEauthorblockA{\IEEEauthorrefmark{2}Charmworks, Inc., Urbana, Illinois, USA\\
	Email: \{zanef2, simengl2, jchoi157, mdiener, kale\}@illinois.edu}}

\maketitle

\begin{abstract}
Python is rapidly becoming the \textit{lingua franca} of machine learning and scientific computing. With the broad use of frameworks such as Numpy, SciPy, and TensorFlow, scientific computing and machine learning are seeing a productivity boost on systems without a requisite loss in performance. While high-performance libraries often provide adequate performance within a node, distributed computing is required to scale Python across nodes and make it genuinely competitive in large-scale high-performance computing. Many frameworks, such as Charm4Py, DaCe, Dask, Legate Numpy, mpi4py, and Ray, scale Python across nodes. However, little is known about these frameworks' relative strengths and weaknesses, leaving practitioners and scientists without enough information about which frameworks are suitable for their requirements. In this paper, we seek to narrow this knowledge gap by studying the relative performance of two such frameworks: Charm4Py and mpi4py.
 
We perform a comparative performance analysis of Charm4Py and mpi4py using CPU and GPU-based microbenchmarks other representative mini-apps for scientific computing.
\end{abstract}

\begin{IEEEkeywords}
performance, analysis, benchmark, HPC, GPU, Python, parallel programming, MPI, Charm++
\end{IEEEkeywords}
\section{Introduction}
Driven by the end of Moore's law, current and future large-scale systems are largely heterogeneous, with different combinations of CPUs, GPUs, and other accelerators such as FPGAs. This increasing heterogeneity is associated with a concordant increase in programming complexity. Recently, the community has focused on performance portability, with frameworks such as Kokkos~\cite{Carter2013:Kokkos} and RAJA~\cite{Beckingsale2019:Raja} providing promises of ``write once, run anywhere''. 

While these systems provide performance portability, we believe they leave much to be desired concerning productivity. Python, the king of productive programming, has for many years been out of the realm of performance-oriented programming: the interpreted language is much too slow to run serious scientific computing workloads on its own. However, a proliferation of performance-oriented Python libraries has recently brought lots of attention in the HPC community. Most notably, the Numpy~\cite{Harris2020:Array} project brings array programming to Python near the speed of raw \texttt{C/C++} code for some applications. 

Numpy demonstrated that Python and high-performance are not orthogonal. Following its success, many other projects accelerate Python, either through bindings to existing C libraries such as CuPy~\cite{Okuta2017:CuPy}, PyOpenCL~\cite{Kloeckner2012:Pycuda}, and PyKokkos~\cite{Awar2021:Performance} or by compiling a subset of Python to native code. Examples of the latter approach are Numba~\cite{Lam2015:Numba}, Pythran~\cite{Guelton2015:Pythran}, or Data-Centric Python~\cite{Ziogas2021:Productivity}.

While these frameworks are well-suited for high-performance within a node, distributed-memory computing is necessary to scale Python beyond one node, making it suitable for high-performance computing. Over the years, different frameworks such as Charm4Py~\cite{Galvez2018:CharmPy}, Dask~\cite{Rocklin2015:Dask}, Legate NumPy~\cite{Bauer2019:Legate}, MPI for Python (mpi4py)~\cite{Dalcin2021:mpi4py}, and Ray~\cite{Moritz2018:Ray} intend in one way or another to fill this performance gap.

Despite the many different applications for high-performance distributed Python, little is known about these frameworks' relative strengths and weaknesses, leaving scientists and practitioners to guess which is best suited for their purposes. In this paper, we seek to narrow this gap by performing a study comparing the performance of two such frameworks: \charmpy{} and \mpipy{}. We choose \charmpy{} and \mpipy{} because each provides Python bindings to widely-used parallel programming models, enabling us to evaluate the performance overhead of Python.

In this paper, we perform a comprehensive performance analysis between \charmpy{} and \mpipy{}. We compare them along different dimensions, comparing microbenchmark performance and performance in representative proxy applications including Stencil2D and a Particle-in-Cell (PIC) code with load imbalance. We perform both CPU and GPU-based benchmarks, discovering the strengths and weaknesses of each framework. 

The rest of this paper is organized as follows:
\begin{enumerate}
    \item Section~\ref{sec:bg} introduces \charmpy{} and \mpipy{}.
    \item Section~\ref{sec:bm_suite} describes in detail our chosen benchmarks.
    \item Section~\ref{sec:perf_eval} contains our performance evaluation and benchmarking results.
    \item Section~\ref{sec:related} performs a survey of related work.
    \item Section~\ref{sec:conclusion} concludes our paper.
\end{enumerate}

\section{Background}\label{sec:bg}
\subsection{Charm4Py}\label{bg_charm4py}
\charmpy{}~\cite{charmpy} is a parallel programming model built on top of \charm{}. \charmpy{} features the message-driven scheduling of \charm{}~\cite{Acun14:Parallel}, and has support for many \charm{} features such as dynamic load balancing, GPU-direct communication~\cite{Choi2021:GPU}, overdecomposition, and sections. Following the programming model of \charm{}, \charmpy{} programs consist of one or more chares on each PE in the computation. These chares communicate through entry methods in a message-driven fashion: chares execute when they have received a message via an entry method invocation.

\subsection{mpi4py}\label{bg_mpi4py}
MPI for Python (mpi4py)~\cite{Dalcin2021:mpi4py} is a popular package providing python bindings for \mpi{} that has been in development since 2005.
MPI for Python uses Cython~\cite{Behnel2011:Cython} as a high-performance middleware between Python and \texttt{C/C++}.

\subsection{Messaging in Python}\label{sub:bg_messaging}
Both \charmpy{} and MPI for Python can send arbitrary Python objects over the network using Python's Pickle framework. Additionally, both have optimizations for objects that implement the buffer protocol, such as Numpy~\cite{Harris2020:Array} arrays. This optimization avoids pickling and allows the underlying buffer to be sent directly over the network.

In addition to optimizations for host-resident data, \charmpy{} and \mpipy{} are capable of inter-process communication consisting of GPU-resident data without first staging data on the host. \charmpy{} uses the underlying UCX capabalities of \charm{}~\cite{Choi2021:GPU}, and \mpipy{} utilizes CUDA-aware \mpi{} implementations.

In the following section, we describe the microbenchmarks and proxy applications presented in this paper.

\section{Benchmark Suite}\label{sec:bm_suite}
\subsection{Communication Microbenchmarks}
We assess the communication performance of both \charmpy{} and \mpipy{} through point-to-point bandwidth and latency CPU and GPU benchmarks using the OSU microbenchmark suite~\cite{Bureddy2012:OMBGPU}. These benchmarks have a twofold purpose: to determine the relative communication performance of \charmpy{} and \mpipy{}, and to evaluate the Python overhead of each. High overhead may result in degraded application performance, potentially rendering a framework unsuitable for large-scale execution. Furthermore, we compare the microbenchmark performance of each framework to their \texttt{C/C++} counterparts. 

On the CPU and GPU, we measure inter-socket and inter-node latency and bandwidth. We refer the reader to~\cite{Bureddy2012:OMBGPU} for a complete description of the benchmarks. 

\subsection{Proxy Application: Jacobi2D}\label{sub:stencil_impl}
Because stencil communication patterns are common in scientific computing workloads, we demonstrate the performance of \charmpy{} and \mpipy{} using the Jacobi iteration in two dimensions. The problem domain is decomposed into rectangular blocks of equal size such that the surface-to-volume ratio is minimized. We assign portions of the domain to PEs as follows.

For $P$ PEs and an $n\times m$ problem domain, we arrange the PEs into a $p_1\times p_2$ rectangular grid, where $p_1\cdot p_2=P$. Each PE is assigned equal portions of the domain of size $(n/p_1)\times (m/p_2)$. Without loss of generality, we assume that $n\mod p_1=0$ and $m\mod p_2=0$. Every iteration, each PE performs the stencil computation and exchanges halo regions with its neighbors. We consider non-periodic boundaries, but do not expect that periodic boundaries conditions will affect our conclusions. 

We use Numpy arrays to represent the problem domain, and Numba is used to JIT-compile computational kernels.

\subsection{Proxy Application: Particle-In-Cell}\label{sub:pic_impl}
To assess the performance of each framework in applications with load imbalance, we implement the particle-in-cell (PIC) parallel research kernel introduced by Georgana~et.~al.~\cite{Georganas2016:Design}. A brief description of the kernel follows. We refer the reader to~\cite{Georganas2016:Design} for a complete description.

The simulation domain is modeled as an $L\times L$ mesh composed of $h\times h$ cells with periodic boundaries in both $x$ and $y$ dimensions. 
%
%
At each timestep, a particle's position and velocity are calculated. Following~\cite{Georganas2016:Design}, we initialize the grid with positive charges at columns with mesh points at odd indices and negative charges at columns with mesh points at even indices. In this paper, we consider the geometric particle distribution introduced by~\cite{Georganas2016:Design}.

Similar to Section~\ref{sub:stencil_impl}, we use Numpy arrays to represent particles, and computational kernels are JIT-compiled using Numba.
%
%
%
In contrast to the Stencil2D kernel, the PIC kernel has application-induced load imbalance. Consequently, we employ overdecomposition in the Charm4Py implementation, though this choice does not affect the implementation: the degree of overdecomposition is a tunable runtime parameter subject to a user's preferences. 


\section{Performance Evaluation}\label{sec:perf_eval}

\subsection{Experimental Setup}
\subsubsection{Hardware Platforms}
We perform our evaluation using two different platforms: \stampede{} and \summit{}. A description of each system can be found in Appendix~\ref{AD}.



\subsubsection{Software Configuration}\label{subsub:config}
In the interest of space, we refer the reader to Appendix~\ref{AD} for a description of the software configuration used in the experiments that follow.




\subsubsection{Evaluation Methodology}
Unless otherwise stated, we perform each experiment for ten trials. The independent variable groups experiments, i.e., all ten trials for an experiment with 768 PEs are in one group. Within a group of 10 trials, the same nodes are used. Between groups, we do not control whether jobs are submitted to the same sets of nodes. Because different calls to \texttt{mpirun} may not be independent, we follow the recommendation of~\cite{Hunold2016:Reproducible} and randomize the order of calls to \texttt{mpirun}\footnote{Or whichever launcher is appropriate for a given platform} within a group of trials. 

Timing begins and ends with a global barrier. Unless otherwise stated, within each trial, ten warmup iterations are performed before timing begins. We report the mean of times recorded from the ten trials and calculate confidence intervals using bootstrapping. We calculate p-values using the non-parametric Mann-Whitney U Test~\cite{Mann1947:On}.
\subsection{Communication Microbenchmarks}\label{sub:comm_microb}
\begin{figure}[t!]
  \begin{subfigure}[b]{.22\textwidth}
    \centering
    \includegraphics[width=\linewidth]{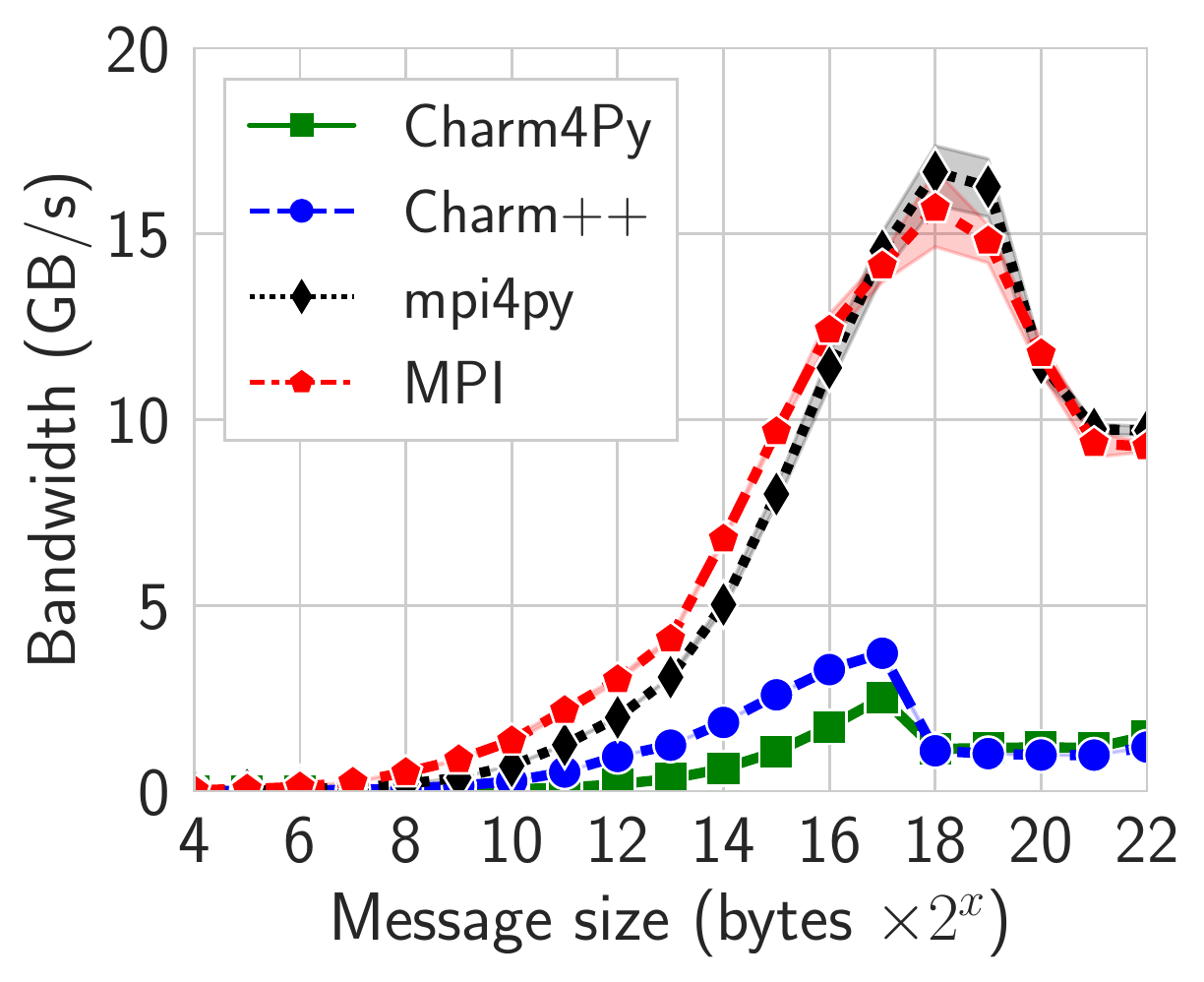}
    \caption{}
  \end{subfigure}
  \hfill
  \begin{subfigure}[b]{.22\textwidth}
    \centering
    \includegraphics[width=\linewidth]{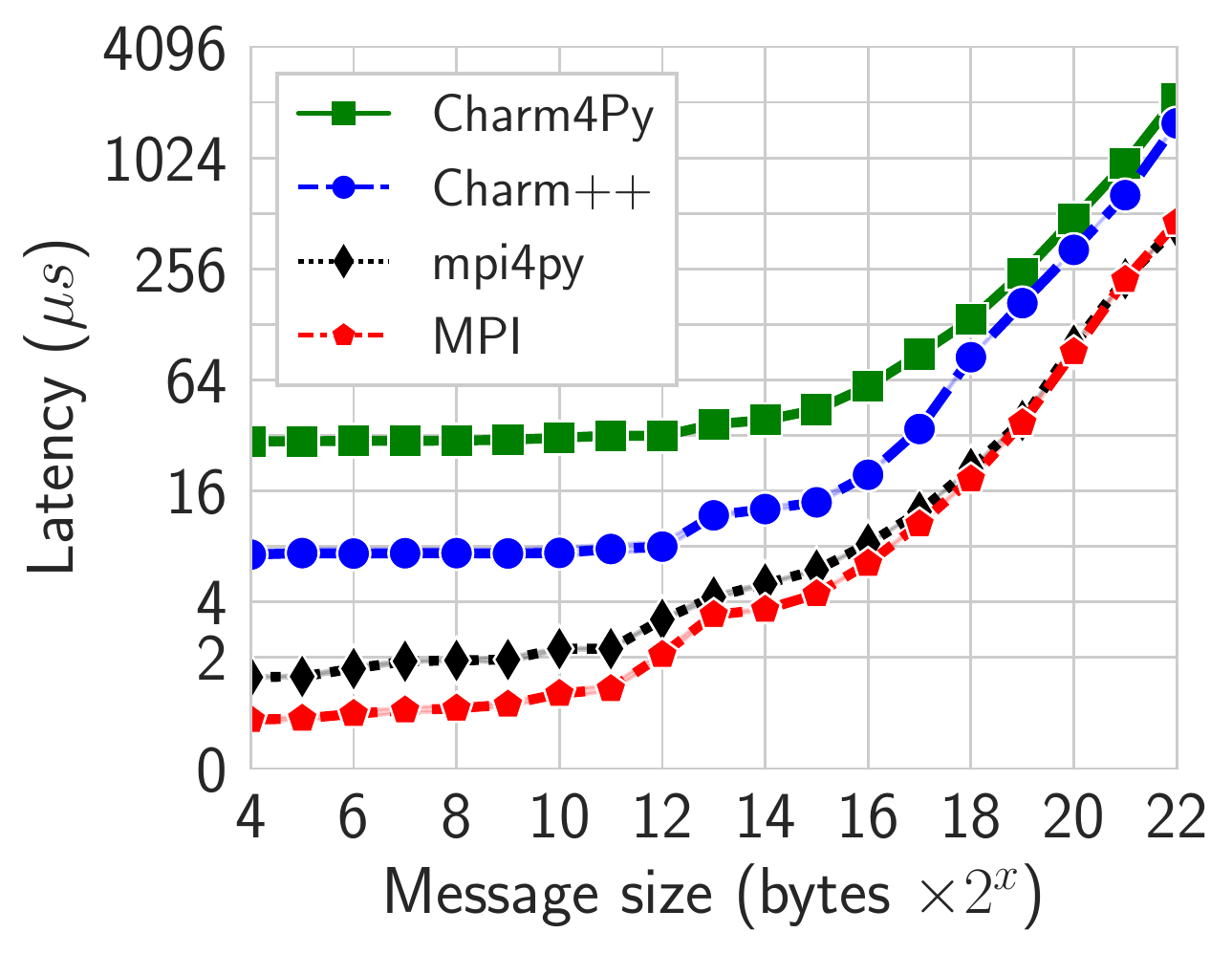}
    \caption{}
  \end{subfigure}
  \begin{subfigure}[b]{.22\textwidth}
    \centering
    \includegraphics[width=\linewidth]{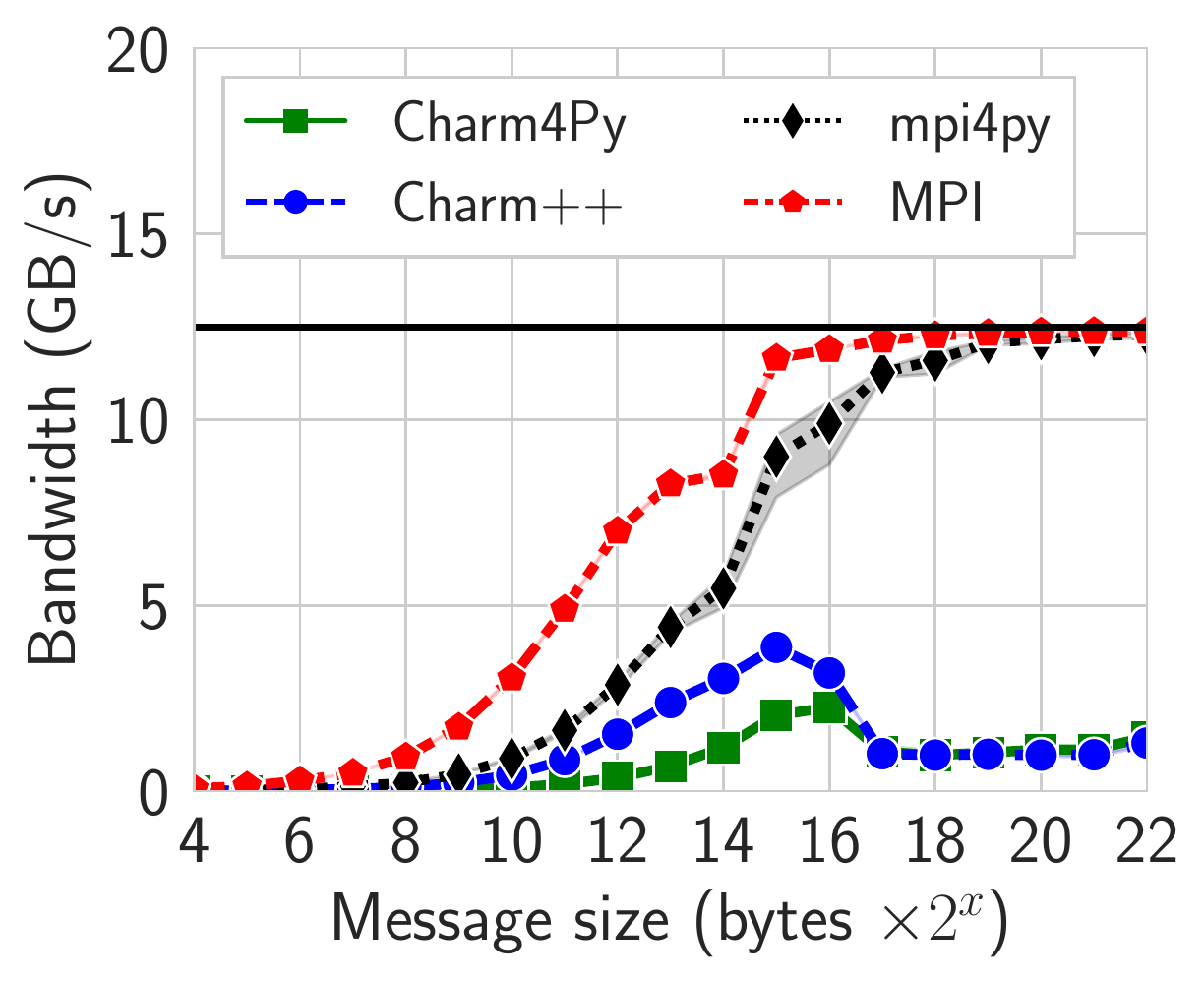}
    \caption{}
  \end{subfigure}
  \hfill
  \begin{subfigure}[b]{.22\textwidth}
    \centering
    \includegraphics[width=\linewidth]{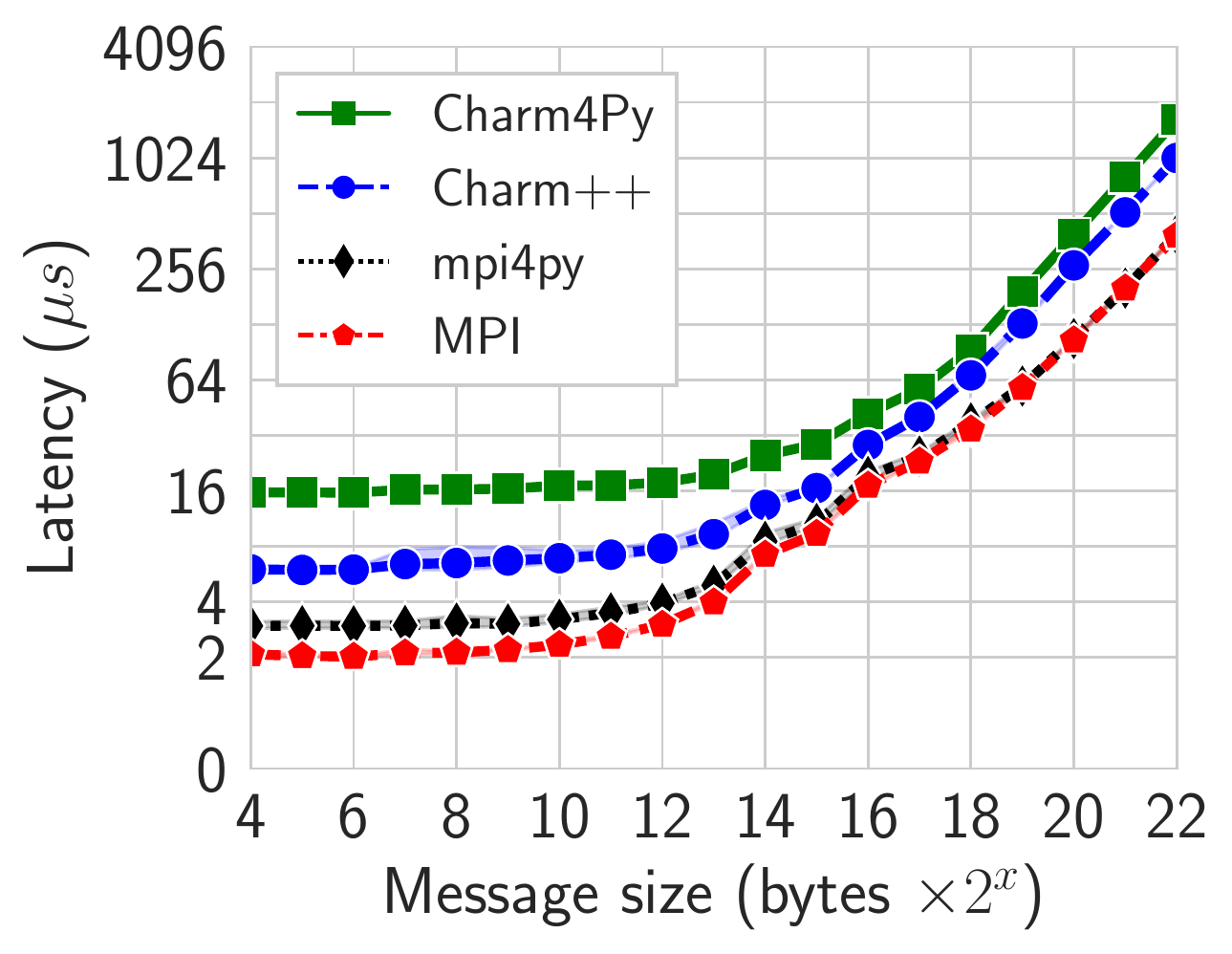}
    \caption{}
  \end{subfigure}
  \caption{Intra-node (inter-socket) and inter-node bandwidth (a), (c) and latency (b), (d) for \charm{}, \charmpy{}, \mpi{}, and \mpipy{}. Message sizes range from $16$ bytes to $4$ MiB. In (c), the horizontal line marks the reported bandwidth of \stampede{}'s interconnection network, 12.5~GB/s.}
  \label{fig:cpu_bw_latency}
\end{figure}
In the microbenchmark experiments that follow, within a trial, 1000 iterations are performed for messages less than 8192 bytes; for messages 8192 bytes or larger, we perform 100 trials. In the latency experiments, we performed 60 warmup iterations; we observed that ten warmup iterations were sufficient for the bandwidth experiments. On \summit{}, we found that setting the rendezvous threshold UCX uses to 131072 bytes yields the best performance for intra-node messages; for inter-node messages, we find that setting the threshold to 8192 bytes provides the best performance.

In Figure~\ref{fig:cpu_bw_latency}, the results for CPU-only inter-socket and inter-node bandwidth (a), (c), and latency (b), (d) respectively, are shown. Because the purpose of this benchmark is to evaluate the overhead \charmpy{} and \mpipy{} induce on the reference frameworks, we disable the use of RDMA in \charm{}. This is because \charmpy{} does not currently support RDMA for host-resident messages. Instead, message marshaling is used, where RDMA would generally be recommended in this benchmark.

From the intra-node data, we make the following observations. First, the Python layer of \mpipy{} is lightweight, resulting in little overhead on top of \mpi{}. MPI for Python has $5\%$ lower\footnote{We found that when \mpipy{} outperformed \mpi{} the result was insignificant ($p>0.01$).} to $86\%$ higher latency than \mpi{} (Figure~\ref{fig:cpu_bw_latency}(b)). On the other hand, the Python layer of \charmpy{} is heavy: we observe that \charmpy{} has $37-311\%$ higher latency than \charm{}. This is because \charmpy{} performs tasks such as chare management in Python, requiring multiple Python function calls and data accesses for each message, whereas \mpipy{} has a lightweight Python layer. We note that this comparison overestimates \mpipy{} overhead relative to \charmpy{}: it is harder to impose less overhead over a faster baseline. Second, the MPI-based frameworks exhibit much greater bandwidth than the Charm-based frameworks (Figure~\ref{fig:cpu_bw_latency}(a)). Recall from Appendix~\ref{AD} that \charm{} is built upon \mpi{} on \stampede{}. Nevertheless, \charm{} achieves $25\%$ of the bandwidth peak of \mpi{}. This is because of the host-staging that the Charm frameworks perform, requiring additional copies.

Inter-node bandwidth and latency are shown in Figure~\ref{fig:cpu_bw_latency}(c), (d), where we see a similar pattern to that observed within a node: the additional copies performed preclude \charmpy{} from achieving peak bandwidth; both \mpi{} and \mpipy{} saturate the node injection bandwidth.

\begin{figure}[]
  \begin{subfigure}[b]{.22\textwidth}
    \centering
    \includegraphics[width=\linewidth]{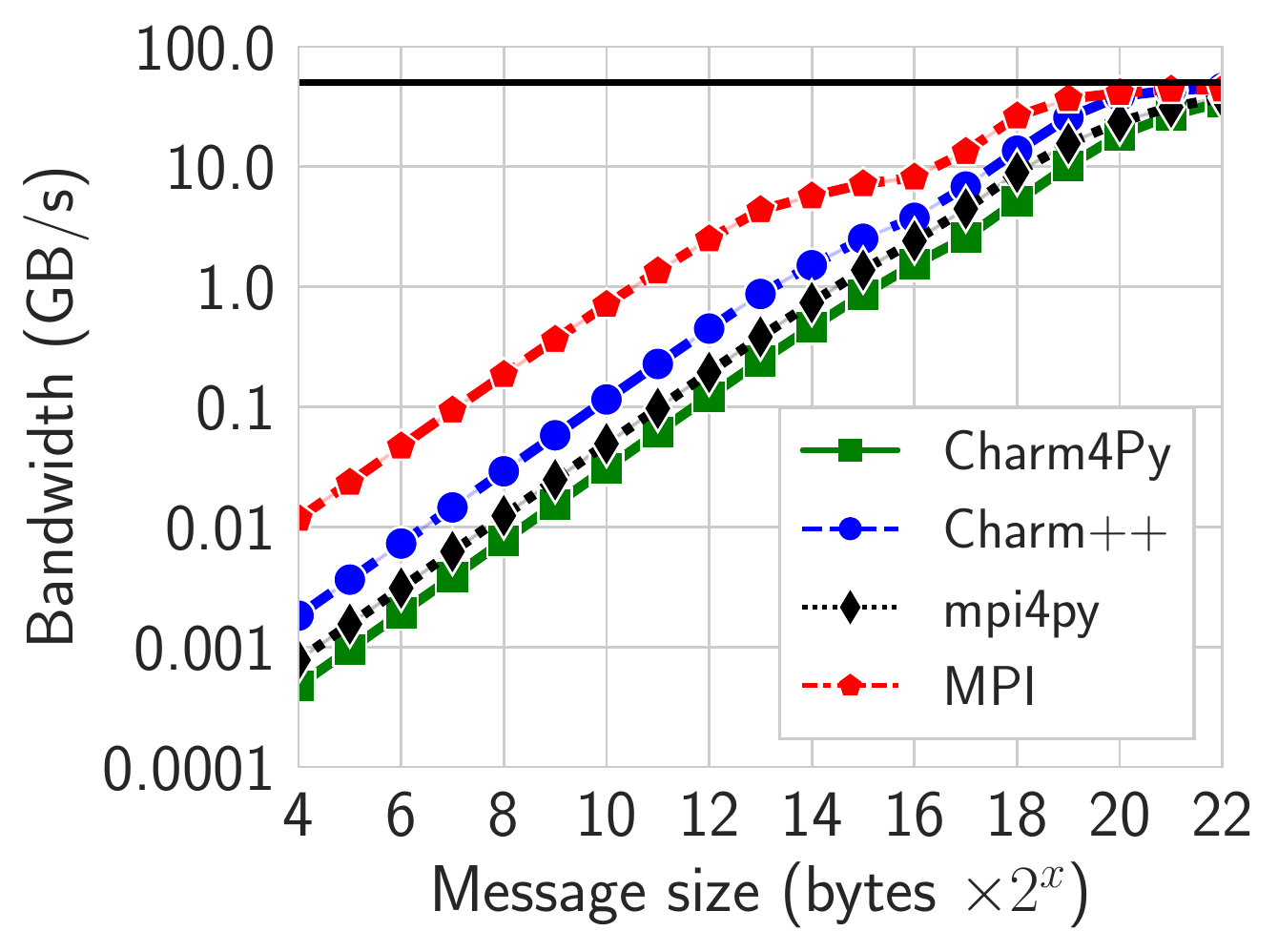}
    \caption{}
  \end{subfigure}
  \hfill
  \begin{subfigure}[b]{.22\textwidth}
    \centering
    \includegraphics[width=\linewidth]{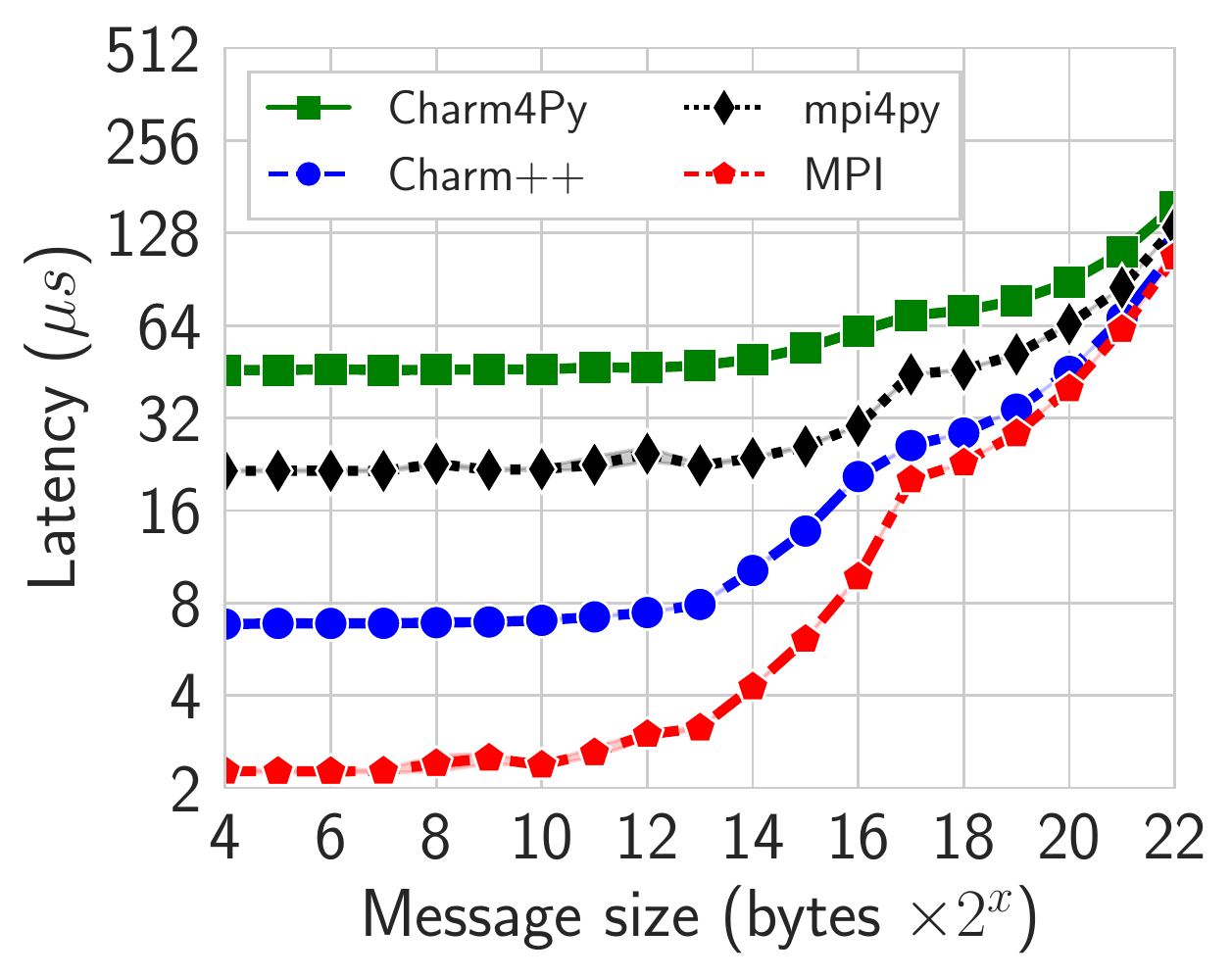}
    \caption{}
    \end{subfigure}
     \begin{subfigure}[b]{.22\textwidth}
    \centering
    \includegraphics[width=\linewidth]{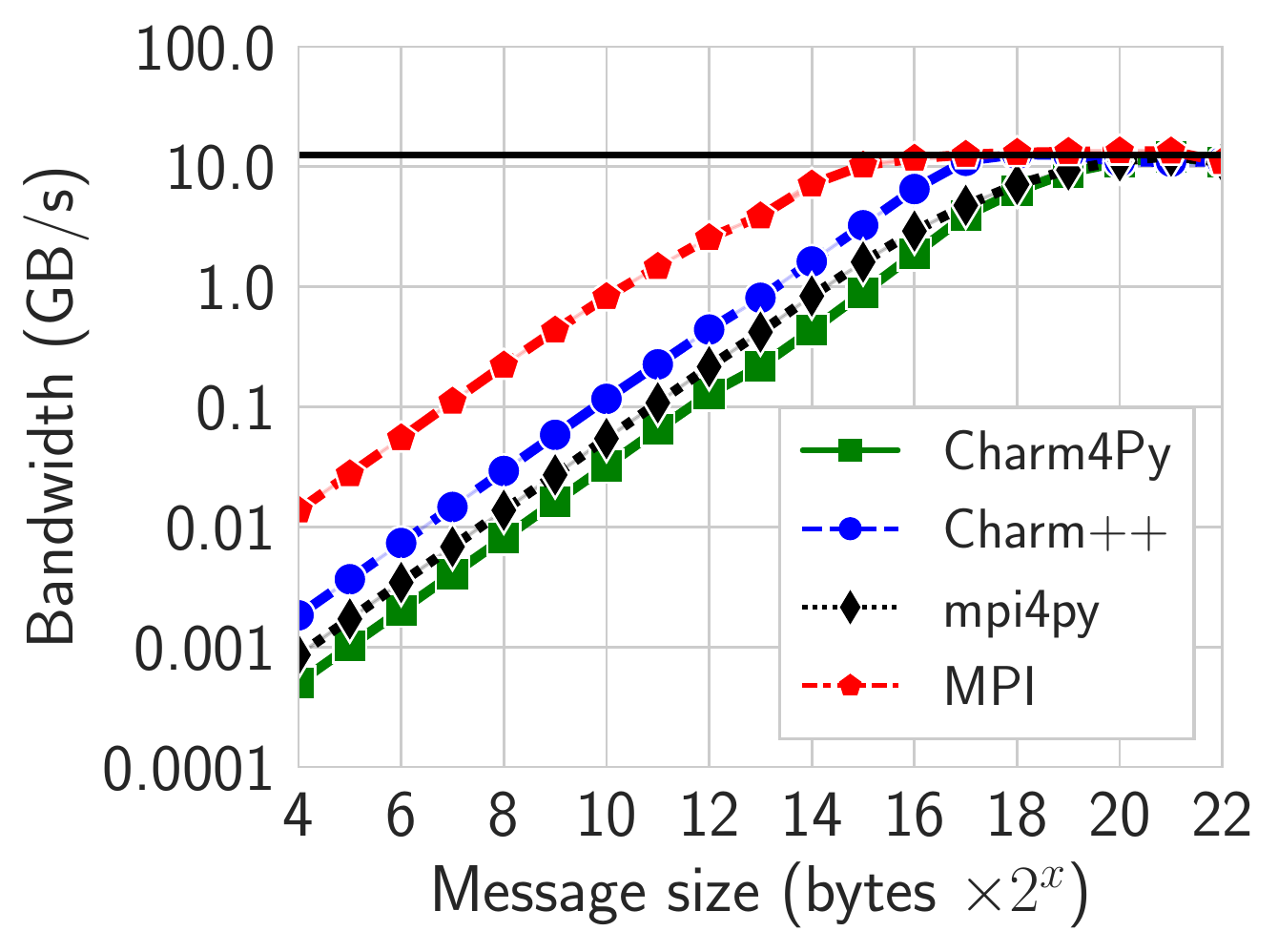}
    \caption{}
  \end{subfigure}
  \hfill
  \begin{subfigure}[b]{.22\textwidth}
    \centering
    \includegraphics[width=\linewidth]{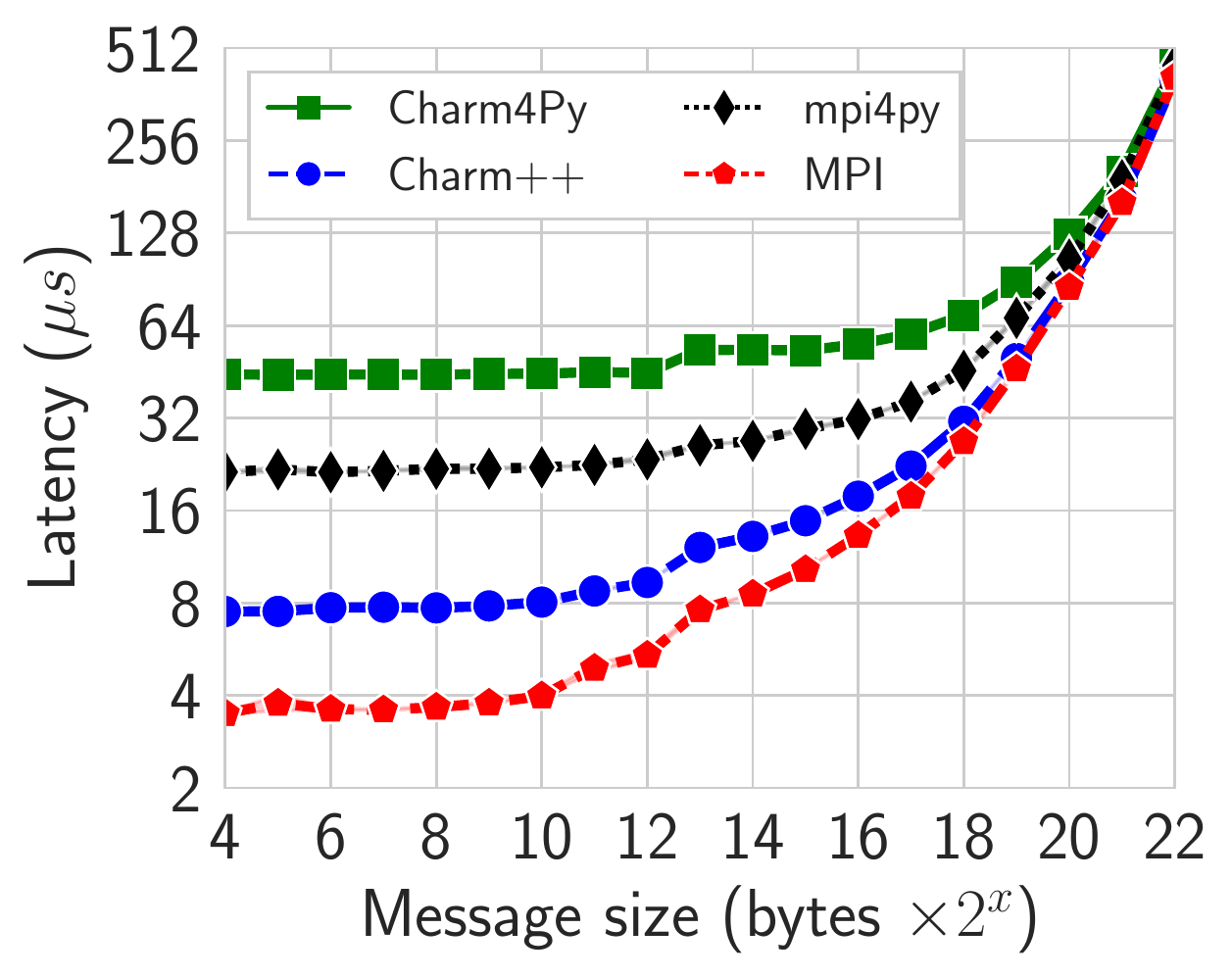}
    \caption{}
  \end{subfigure}
    \caption{Intra-node bandwidth (GB/s) and latency ($\mu s$) (a), (b) and inter-node bandwidth (MB/s) and latency ($\mu s$) (c), (d) for GPU-resident data on \summit{}. In (a), the black horizontal line is at 50 GB/s: the reported bandwidth of the NVLink connection between GPUs within a node. In (c) the horizontal black line is at 12.5GB/s, the stated bandwidth of the interconnection network of \summit{}.}
    \label{fig:gpu_bw_latency}
\end{figure}
\begin{figure}[]
  \begin{subfigure}[b]{.22\textwidth}
    \centering
    \includegraphics[width=\linewidth]{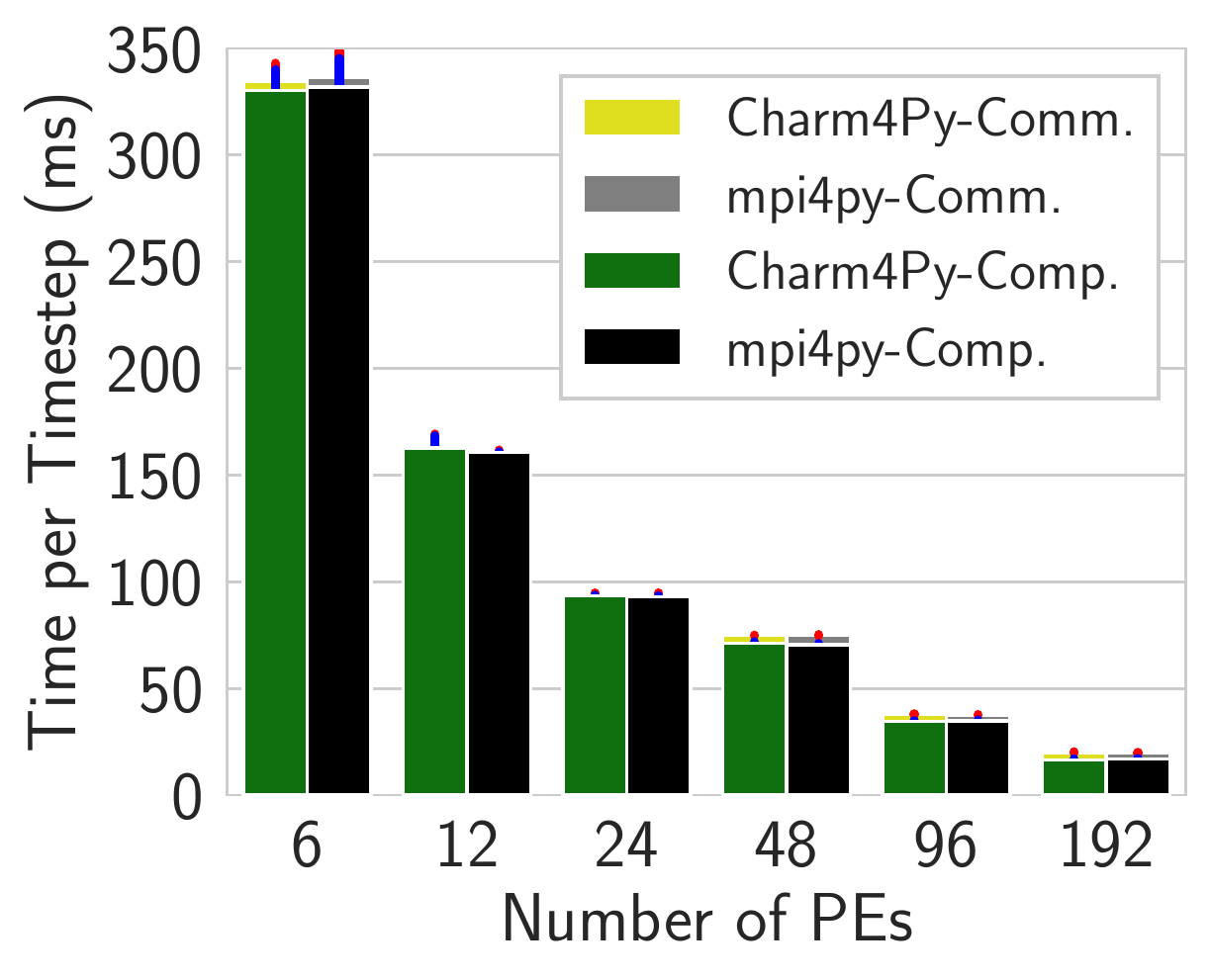}
    \caption{}
  \end{subfigure}
  \hfill
  \begin{subfigure}[b]{.22\textwidth}
    \centering
    \includegraphics[width=\linewidth]{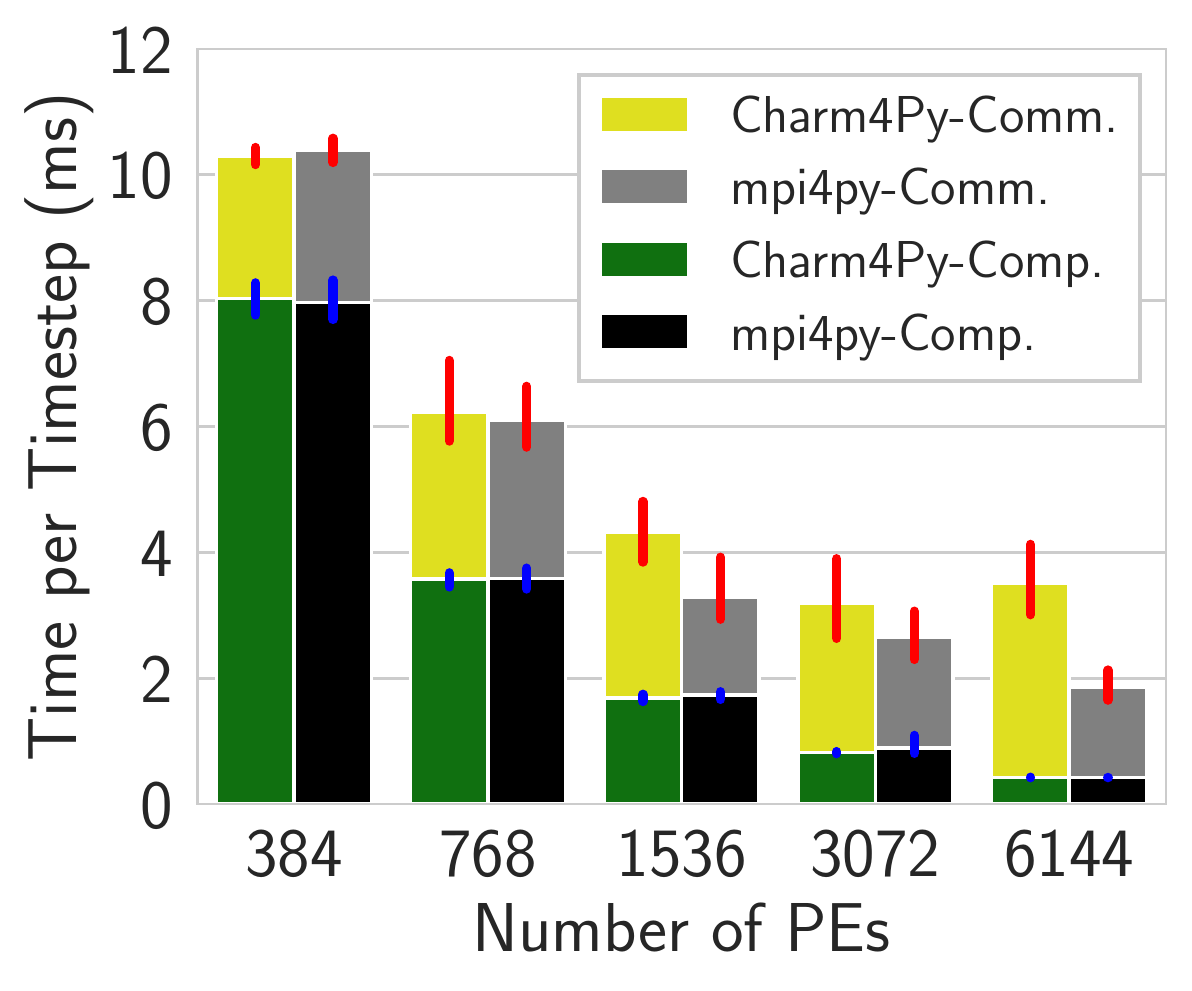}
    \caption{}
  \end{subfigure}
\caption{Time per timestep (in ms) vs. the number of PEs used for Stencil2D on\stampede{}. Results are for strong-scaling. In (a), we show results for PE counts between 6 and 192 (1-4 nodes); in (b), results are shown for PE counts between 384 and 6144 (8-128 nodes). Error bars designate the 99\% CI of the mean for computation time and communication time in blue and red, respectively.}
  \label{fig:stencil_cpu_strong}
\end{figure}
In Figure~\ref{fig:gpu_bw_latency}, the intra-node bandwidth and latency are shown in (a) and (b), and the inter-node bandwidth and latency are shown in (c), (d). Recall from Section~\ref{sub:bg_messaging} that all frameworks use GPU-direct for messages containing GPU-resident data. Contrasting CPU messages, \mpipy{} imposes substantial overhead over the raw \mpi{} calls. We observe that \mpipy{} latency is between $24\%$ and $851\%$ higher than \mpi{}; for \charmpy{} we see latency between $38\%$ and $570\%$ higher than \charm{}. This is because both \charmpy{} and \mpipy{} must extract metadata for the underlying device buffer from the host abstraction for CUDA data. This metadata lookup involves accessing a Python dictionary and several attribute lookups.

In the bandwidth figures, we see that \mpi{} has substantially higher bandwidth than all other frameworks. However, we find that for both intra-node messages Figure~\ref{fig:gpu_bw_latency}(a) and inter-node messages Figure~\ref{fig:gpu_bw_latency}(b), all frameworks eventually saturate the bandwidth of the link being used.

\subsubsection{Stencil2D}
In Figure~\ref{fig:stencil_cpu_strong}, we see the strong-scaling results for the CPU implementation of the Stencil2D proxy application. The problem domain is $24576\times 24576$. Recall that each node of \stampede{} has 48 total cores, 24 per socket. We find that below 768 PEs, the differences between \charmpy{} and \mpipy{} are insignificant ($p>0.01$). However, beyond 768 PEs, the \charmpy{} implementation suffers. Specifically, we can see that the communication performance observed in Section~\ref{sub:comm_microb} results in degraded communication performance for \charmpy{}, limiting scalability.

\begin{figure}[t]
\centering
\includegraphics[width=0.3\textwidth]{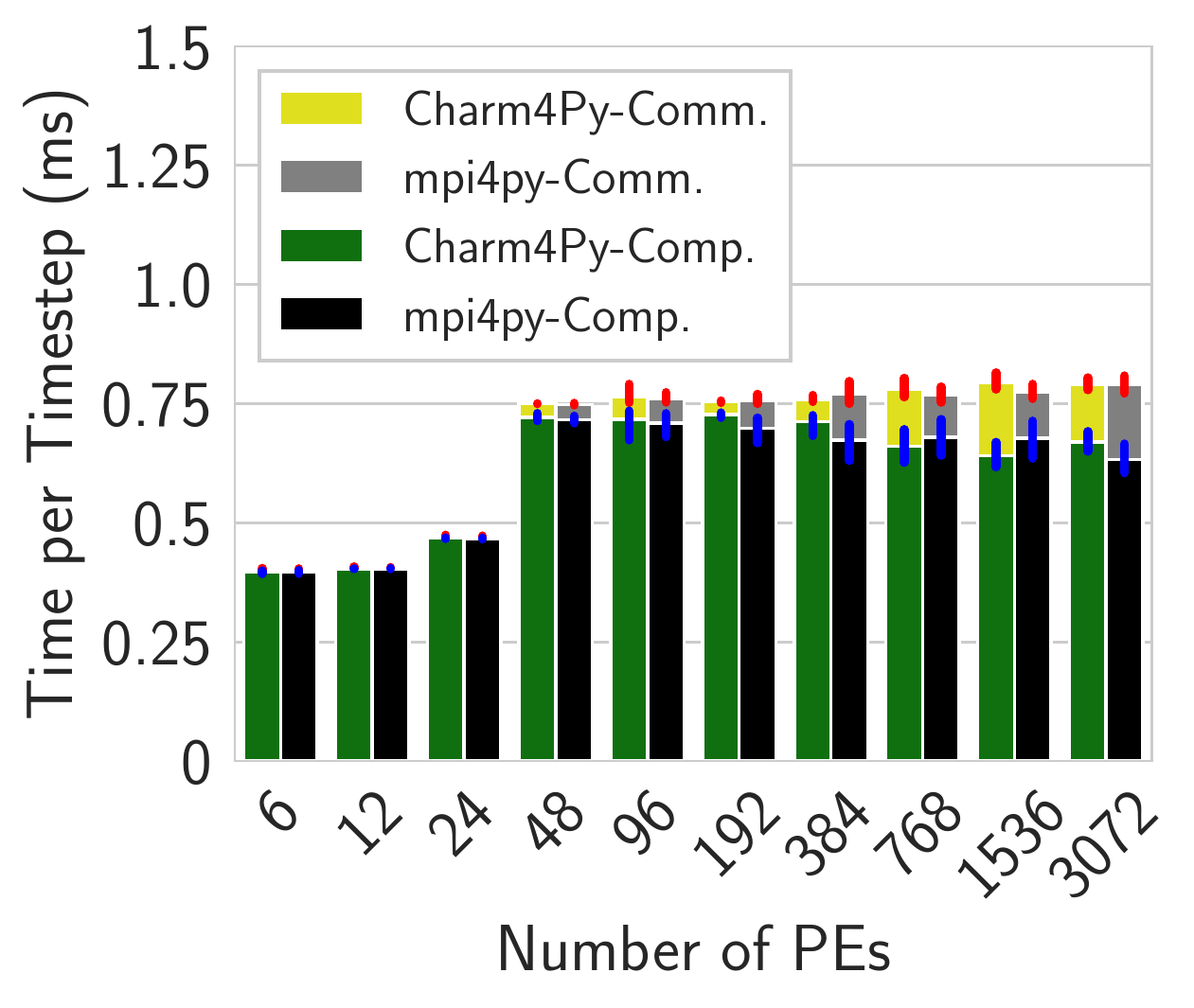}
\caption{Time per timestep (ms) vs. the number of PEs used in the Stencil2D proxy application weak-scaled on \stampede{}.} 
\label{fig:stencil_cpu_weak}
\end{figure}

In Figure~\ref{fig:stencil_cpu_weak}, weak scaling results for the Stencil2D application are shown. To weak scale the problem, we begin with a domain of $6144\times 6144$. We first double the domain in the x-dimension and then the y-dimension. We observe that both implementations weak scale well from 48-3072 cores. We are investigating the increase in time between 24 and 48 PEs, though we suspect resource contention as the simulation grows from utilizing one to two sockets.

\begin{figure}[t]
  \begin{subfigure}[b]{.22\textwidth}
    \centering
    \includegraphics[width=\linewidth]{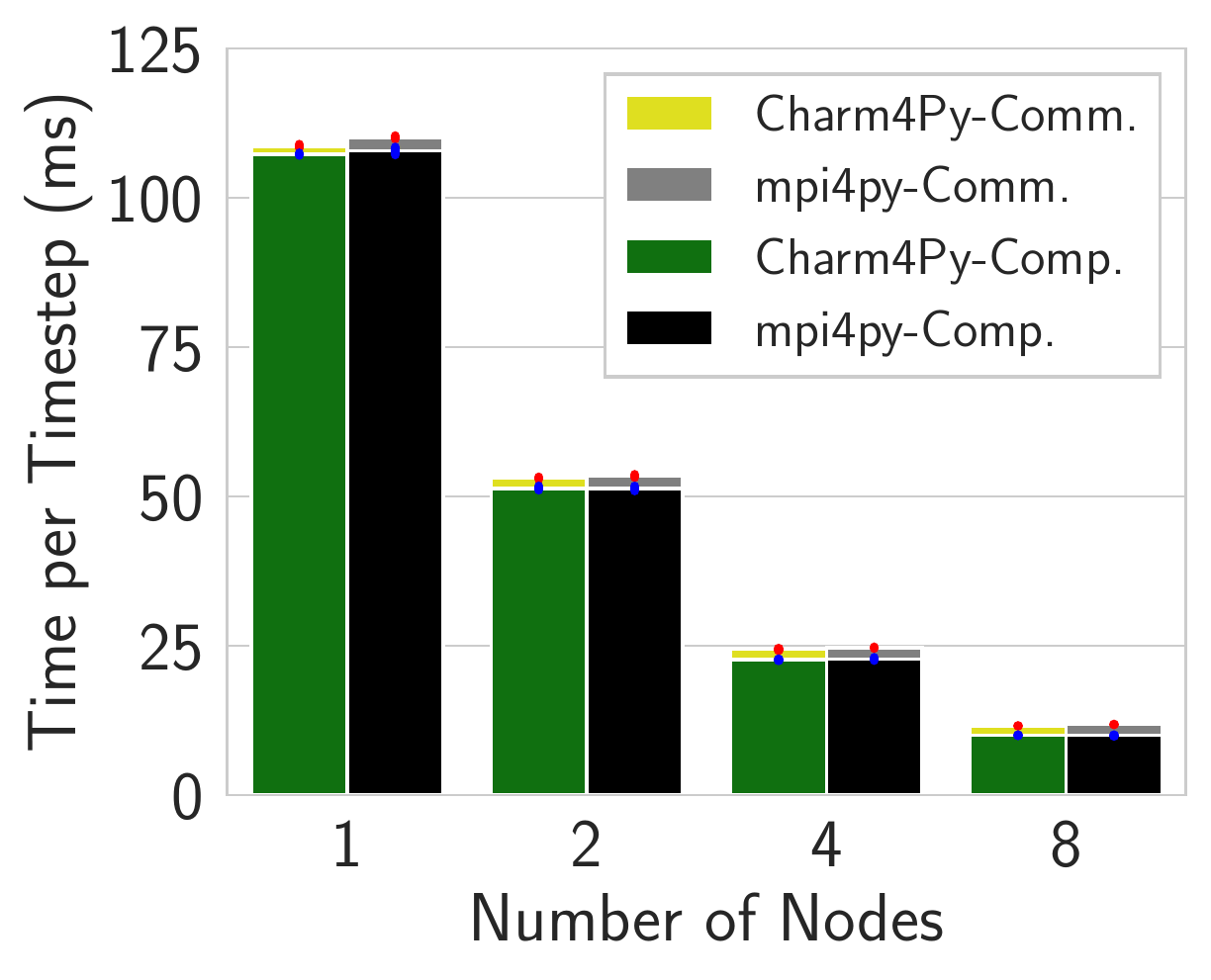}
    \caption{}
  \end{subfigure}
  \hfill
  \begin{subfigure}[b]{.22\textwidth}
    \centering
    \includegraphics[width=\linewidth]{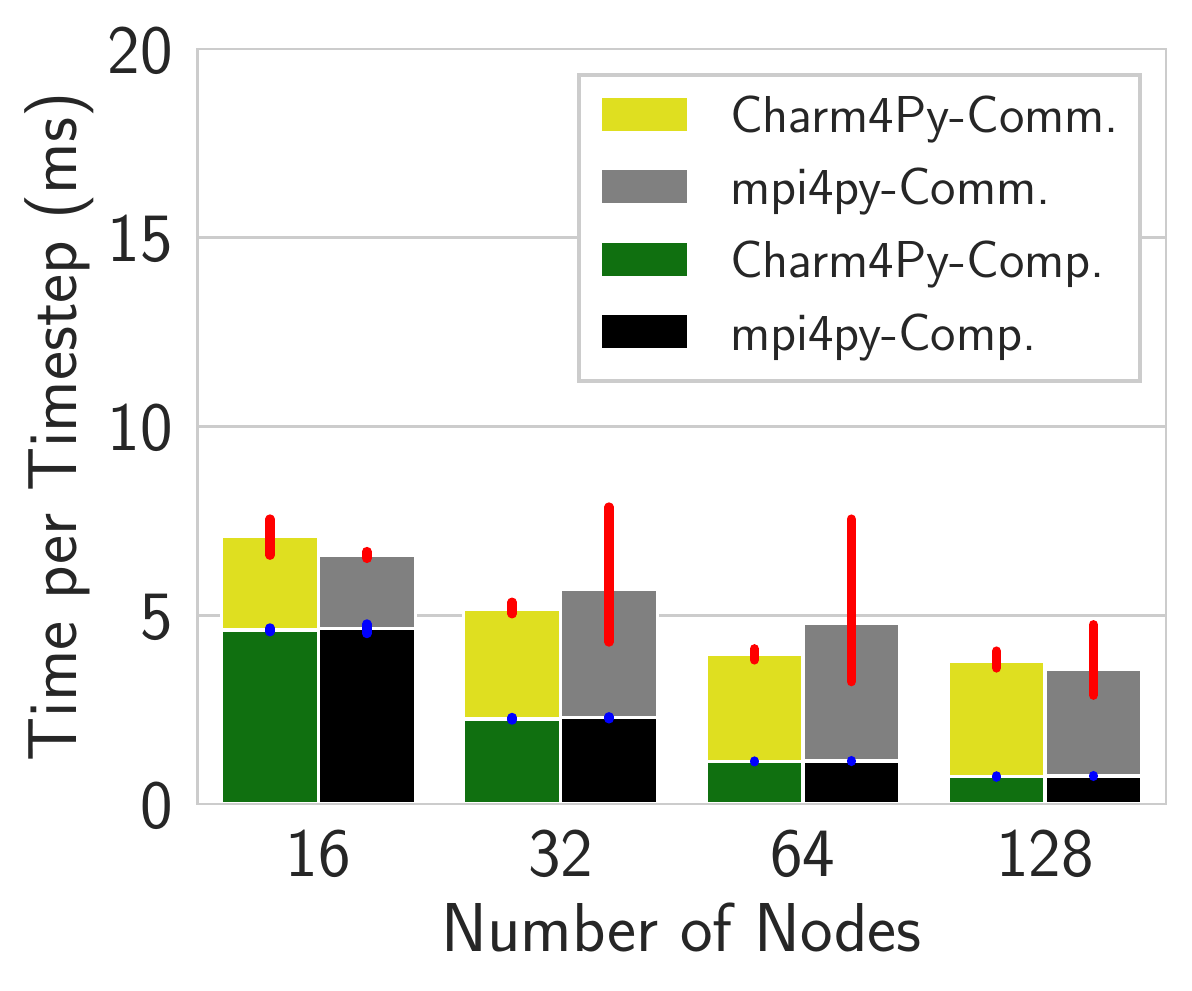}
    \caption{}
  \end{subfigure}
\caption{Time per timestep (in ms) vs. the number of nodes used in the Stencil2D proxy application on strong-scaled \summit{}. In (a), we show results for small node counts (6-48 GPUs); (b) contains results at larger node counts (72-768 GPUs). Each node is equipped with 6 GPUs. The 99\% CI of the mean for computation and communication are outlined in the error bows in blue and red, respectively.}
\label{fig:gpu_stencil_strong}
\end{figure}

Figure~\ref{fig:gpu_stencil_strong} shows the strong-scaling performance of the GPU implementation of Stencil2D. In this experiment, the domain is $73728\times 73728$, approximately the largest domain size that fits in the available 96GiB of device memory on one node. Consistent with the microbenchmark results in Section~\ref{sec:perf_eval}, we find that \charmpy{} and \mpipy{} perform similarly. Indeed, no significant difference in their performance is observed.

\begin{figure}[t]
\centering
\includegraphics[width=0.3\textwidth]{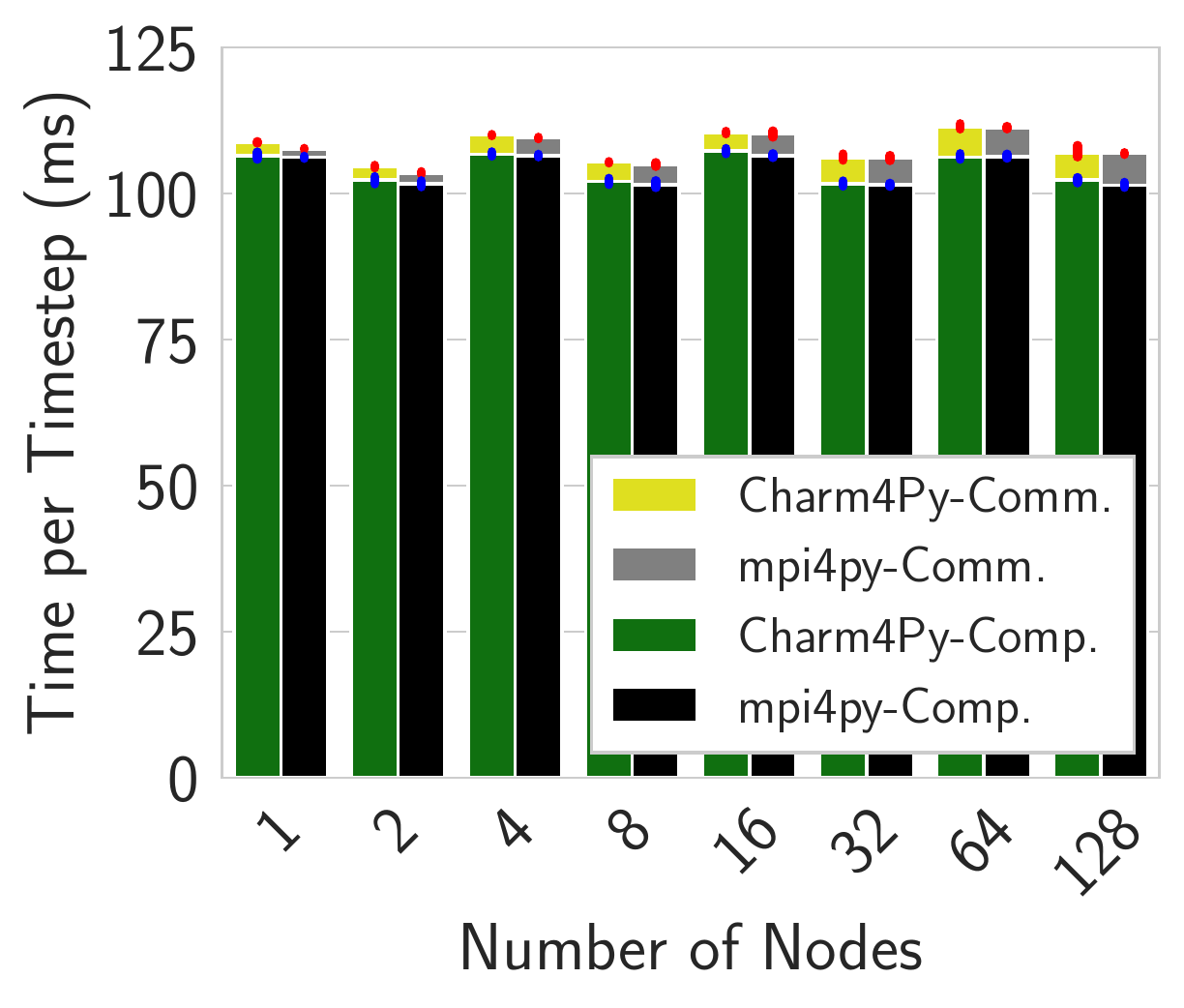}
\caption{Time per timestep (in ms) vs. the number of nodes used in the simulation for weak scaling performance data for Stencil2D on \summit{}.}
\label{fig:gpu_scaling_weak}
\end{figure}

In Figure~\ref{fig:gpu_scaling_weak}, we observe the weak scaling performance of \charmpy{} and \mpipy{}. The beginning problem domain is $73728\times 73728$. We scale first in the x-dimension and then in the y-dimension. This figure shows that both implementations weak-scale well (parallel efficiency $E\approx 1$) and exhibit similar communication performance characteristics.

\subsubsection{Particle-In-Cell}
\begin{figure}[t]
    \centering
    \includegraphics[width=0.3\textwidth]{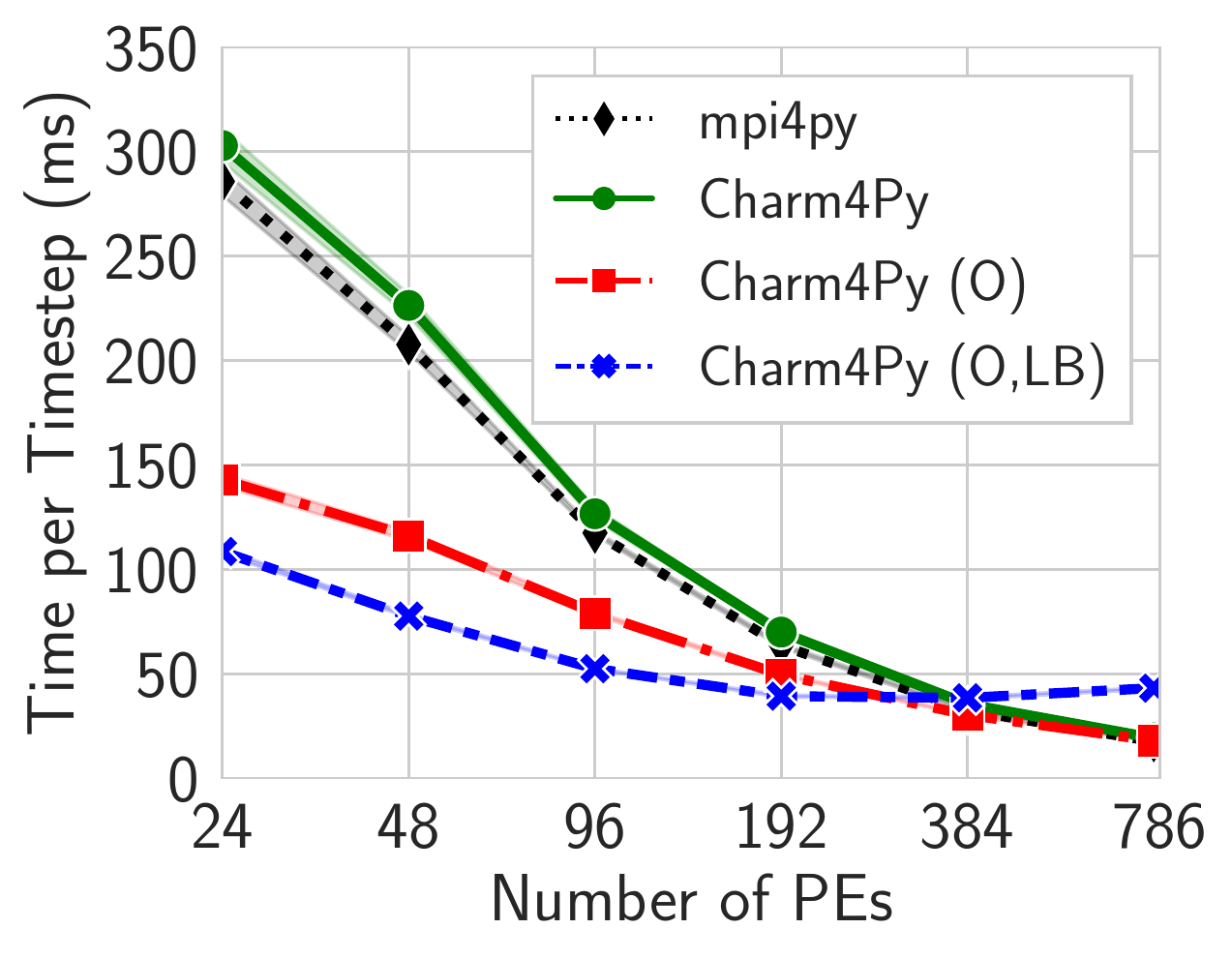}
    \caption{Time per timestep (in milliseconds) vs. the number of PEs used for \charmpy{} and \mpipy{} for the particle-in-cell research kernel executed on\stampede{}. Charm4Py is shown with overdecomposition only (O) and overdecomposition and load balancing (O, LB). Shaded areas outline the $99\%$ CI of the mean.}
    \label{fig:pic_strong_scaling}
\end{figure}

To evaluate the performance of \charmpy{} and \mpipy{} in the PIC kernel, we use a grid of size $2998\times 2998$ with $600,000$ particles. The simulation is run for $1000$ iterations. Following Georganas et.al.~\cite{Georganas2016:Design}, we use an exponential distribution of particles with $r=0.999$, and $k=0$. We find that \charmpy{} performs best with an overdecomposition factor (ODF) of 8 chares per PE, and that performance is best when load balancing is performed every 80 iterations.

The results of the simulation are shown in Figure~\ref{fig:pic_strong_scaling}. We find that without overdecomposition or load balancing, Charm4Py is out-performed by \mpipy{} up to 96 PEs. The simulation features a high communication volume with few processors, as many particles migrate from one PE to another. However, we see that overdecomposition makes a substantial impact on \charmpy{} performance. This is because overdecomposition results in a different assignment of chares to processors, yielding a more balanced computation. Explicit load balancing improves \charmpy{} performance even further: a speedup up to $2.6\times$ is observed over \mpipy{}. At large PE counts, the cost of migrating chares to balance load is high, as chares have to be serialized and deserialized using expensive pickling operations. We are investigating methods to reduce the cost of chare migration in \charmpy{}.
\section{Related Work}\label{sec:related}
A substantial body of work comparing parallel programming models exists. In~\cite{Karlin2013:Exploring} the LULESH mini-app is used to investigate the performance and productivity aspects of parallel programming models. They consider only the LULESH mini-app, and Python is not considered. Slaughter et al. ~\cite{Slaughter2020:Task} present Task Bench, a parameterized benchmarking platform for programming models. In their evaluation, they only have coverage of Dask in the Python programming model space. The parallel research kernels~\cite{Wijngaart2014:Parallel} are a project that covers many different kernels in many different programming models. However, these kernels lack Python coverage. NPBench~\cite{Ziogas2021:NPBench} is a benchmarking suite for Numpy implementations, but they do not consider distributed computing.
\section{Conclusion}\label{sec:conclusion}
With increasing heterogeneity in current and future large-scale systems, Python's productivity and performance portability make it attractive to scientists and practitioners. In this paper, we have conducted a comprehensive performance analysis of \charmpy{} and \mpipy{} using a set of microbenchmarks and representative mini-apps. We found that in CPU-based applications with uniform load balance between PEs that both \charmpy{} and \mpipy{} perform comparably when the granularity of tasks is large; at smaller task granularities, the poor communication performance of \charmpy{} dominates, and strong-scaling performance is limited. Work to add the RDMA support of \charm{} to \charmpy{} is required to bring the communication performance of \charmpy{} to the level observed in \mpipy{}. Following this implementation, we anticipate comparable CPU-based communication performance between \charmpy{} and \mpipy{}, as we observed in our GPU performance results. In applications with load imbalance between processing elements, we find that \charmpy{} makes effective use of the load-balancing capabilities of \charm{}, yielding speedup up to $2.6\times$ over \mpipy{}.

Future work includes considering a broader set of frameworks and benchmarks to model workloads found in big data analytics and machine learning.
\section*{Acknowledgement}

This material is based in part upon work supported by the Department of Energy, National Nuclear Security Administration, under Award Number \textit{DE-NA0003963}.

This work used the Extreme Science and Engineering Discovery Environment (XSEDE), which is supported by National Science Foundation grant number \textit{ACI-1548562}.

The authors acknowledge the Texas Advanced Computing Center (TACC) at The University of Texas at Austin for providing HPC resources that have contributed to the research results reported within this paper.

This research used resources of the Oak Ridge Leadership Computing Facility, which is a DOE Office of Science User Facility supported under Contract \textit{DE-AC05-00OR22725}.

\printbibliography

@article{Awar2021:Performance,
  title = {A {{Performance Portability Framework}} for {{Python}}},
  author = {Awar, Nader Al and Zhu, Steven and Biros, George and Gligoric, Milos},
  date = {2021},
  pages = {12},
  langid = {english}
}

@inproceedings{Bauer2019:Legate,
  title = {Legate {{NumPy}}: Accelerated and Distributed Array Computing},
  shorttitle = {Legate {{NumPy}}},
  booktitle = {Proceedings of the {{International Conference}} for {{High Performance Computing}}, {{Networking}}, {{Storage}} and {{Analysis}}},
  author = {Bauer, Michael and Garland, Michael},
  date = {2019-11-17},
  pages = {1--23},
  publisher = {{ACM}},
  location = {{Denver Colorado}},
  langid = {english}
}

@incollection{Bureddy2012:OMBGPU,
  title = {{{OMB}}-{{GPU}}: {{A Micro}}-{{Benchmark Suite}} for {{Evaluating MPI Libraries}} on {{GPU Clusters}}},
  booktitle = {Recent {{Advances}} in the {{Message Passing Interface}}},
  author = {Bureddy, D. and Wang, H. and Venkatesh, A. and Potluri, S. and Panda, D. K.},
  date = {2012},
  series = {Lecture {{Notes}} in {{Computer Science}}},
  volume = {7490},
  pages = {110--120},
  langid = {english}
}

@inproceedings{Galvez2018:CharmPy,
  title = {{{CharmPy}}: {{A Python Parallel Programming Model}}},
  shorttitle = {{{CharmPy}}},
  booktitle = {2018 {{IEEE International Conference}} on {{Cluster Computing}} ({{CLUSTER}})},
  author = {Galvez, Juan J. and Senthil, Karthik and Kale, Laxmikant},
  date = {2018-09},
  eventtitle = {2018 {{IEEE International Conference}} on {{Cluster Computing}} ({{CLUSTER}})}
}

@inproceedings{Georganas2016:Design,
  title = {Design and {{Implementation}} of a {{Parallel Research Kernel}} for {{Assessing Dynamic Load}}-{{Balancing Capabilities}}},
  booktitle = {2016 {{IEEE International Parallel}} and {{Distributed Processing Symposium}} ({{IPDPS}})},
  author = {Georganas, Evangelos and Van Der Wijngaart, Rob F. and Mattson, Timothy G.},
  date = {2016-05},
  pages = {73--82},
  issn = {1530-2075},
  eventtitle = {2016 {{IEEE International Parallel}} and {{Distributed Processing Symposium}} ({{IPDPS}})}
}

@article{Harris2020:Array,
  title = {Array Programming with {{NumPy}}},
  author = {Harris, Charles R. and Millman, K. Jarrod and van der Walt, Stéfan J. and Gommers, Ralf and Virtanen, Pauli and Cournapeau, David and Wieser, Eric and Taylor, Julian and Berg, Sebastian and Smith, Nathaniel J. and Kern, Robert and Picus, Matti and Hoyer, Stephan and van Kerkwijk, Marten H. and Brett, Matthew and Haldane, Allan and del Río, Jaime Fernández and Wiebe, Mark and Peterson, Pearu and Gérard-Marchant, Pierre and Sheppard, Kevin and Reddy, Tyler and Weckesser, Warren and Abbasi, Hameer and Gohlke, Christoph and Oliphant, Travis E.},
  options = {useprefix=true},
  date = {2020-09},
  journaltitle = {Nature},
  volume = {585},
  number = {7825},
  pages = {357--362},
  publisher = {{Nature Publishing Group}},
  issue = {7825},
  langid = {english}
}

@online{Moritz2018:Ray,
  title = {Ray: {{A Distributed Framework}} for {{Emerging AI Applications}}},
  shorttitle = {Ray},
  author = {Moritz, Philipp and Nishihara, Robert and Wang, Stephanie and Tumanov, Alexey and Liaw, Richard and Liang, Eric and Elibol, Melih and Yang, Zongheng and Paul, William and Jordan, Michael I. and Stoica, Ion},
  date = {2018-09-29},
  archiveprefix = {arXiv}
}

@inproceedings{Rocklin2015:Dask,
  title = {Dask: {{Parallel Computation}} with {{Blocked}} Algorithms and {{Task Scheduling}}},
  shorttitle = {Dask},
  author = {Rocklin, Matthew},
  date = {2015},
  pages = {126--132},
  location = {{Austin, Texas}},
  eventtitle = {Python in {{Science Conference}}}
}

@inbook{Slaughter2020:Task,
author = {Slaughter, Elliott and Wu, Wei and Fu, Yuankun and Brandenburg, Legend and Garcia, Nicolai and Kautz, Wilhem and Marx, Emily and Morris, Kaleb S. and Cao, Qinglei and Bosilca, George and Mirchandaney, Seema and Lee, Wonchan and Treichler, Sean and McCormick, Patrick and Aiken, Alex},
title = {Task Bench: A Parameterized Benchmark for Evaluating Parallel Runtime Performance},
year = {2020},
publisher = {IEEE Press},
booktitle = {SC20},
articleno = {62},
numpages = {15}
}

@inproceedings{Ziogas2021:NPBench,
  title = {{{NPBench}}: A Benchmarking Suite for High-Performance {{NumPy}}},
  booktitle = {Proceedings of the {{ACM International Conference}} on {{Supercomputing}}},
  author = {Ziogas, Alexandros Nikolaos and Ben-Nun, Tal and Schneider, Timo and Hoefler, Torsten},
  date = {2021-06-03},
  pages = {63--74},
  publisher = {{ACM}},
  eventtitle = {{{ICS}} '21: 2021 {{International Conference}} on {{Supercomputing}}},
  langid = {english}
}

@online{Ziogas2021:Productivity,
  title = {Productivity, {{Portability}}, {{Performance}}: {{Data}}-{{Centric Python}}},
  shorttitle = {Productivity, {{Portability}}, {{Performance}}},
  author = {Ziogas, Alexandros Nikolaos and Schneider, Timo and Ben-Nun, Tal and Calotoiu, Alexandru and De Matteis, Tiziano and Licht, Johannes de Fine and Lavarini, Luca and Hoefler, Torsten},
  date = {2021-07-01},
  archiveprefix = {arXiv}
}

@INPROCEEDINGS {Choi2021:GPU,
author = {J. Choi and Z. Fink and S. White and N. Bhat and D. F. Richards and L. V. Kale},
booktitle = {2021 IEEE International Parallel and Distributed Processing Symposium Workshops (IPDPSW)},
title = {GPU-aware Communication with UCX in Parallel Programming Models: Charm++, MPI, and Python},
year = {2021},
volume = {},
issn = {},
pages = {479-488},
address = {Los Alamitos, CA, USA},
month = {jun}
}

@inproceedings{Lam2015:Numba,
author = {Lam, Siu Kwan and Pitrou, Antoine and Seibert, Stanley},
title = {Numba: A LLVM-Based Python JIT Compiler},
year = {2015},
address = {New York, NY, USA},
doi = {10.1145/2833157.2833162},
booktitle = {Proceedings of the Second Workshop on the LLVM Compiler Infrastructure in HPC},
articleno = {7},
numpages = {6},
location = {Austin, Texas},
series = {LLVM '15}
}

@INPROCEEDINGS{charmpy,
  author={Galvez, Juan J. and Senthil, Karthik and Kale, Laxmikant},
  booktitle={2018 IEEE International Conference on Cluster Computing (CLUSTER)}, 
  title={CharmPy: A Python Parallel Programming Model}, 
  year={2018},
  volume={},
  number={},
  pages={423-433},
  doi={10.1109/CLUSTER.2018.00059}}

@ARTICLE{Dalcin2021:mpi4py,
  author={Dalcin, Lisandro and Fang, Yao-Lung L.},
  journal={Computing in Science   Engineering}, 
  title={mpi4py: Status Update After 12 Years of Development}, 
  year={2021},
  volume={23},
  number={4},
  pages={47-54},
  doi={10.1109/MCSE.2021.3083216}}

@ARTICLE{Behnel2011:Cython,
  author={Behnel, Stefan and Bradshaw, Robert and Citro, Craig and Dalcin, Lisandro and Seljebotn, Dag Sverre and Smith, Kurt},
  journal={Computing in Science   Engineering}, 
  title={Cython: The Best of Both Worlds}, 
  year={2011},
  volume={13},
  number={2},
  pages={31-39},
  }

@INPROCEEDINGS{Carter2013:Kokkos,
  author={Edwards, H. Carter and Trott, Christian R.},
  booktitle={2013 Extreme Scaling Workshop (xsw 2013)}, 
  title={Kokkos: Enabling Performance Portability Across Manycore Architectures}, 
  year={2013},
  volume={},
  number={},
  pages={18-24},
  doi={10.1109/XSW.2013.7}}

@INPROCEEDINGS{Beckingsale2019:Raja,
  author={Beckingsale, David A. and Burmark, Jason and Hornung, Rich and Jones, Holger and Killian, William and Kunen, Adam J. and Pearce, Olga and Robinson, Peter and Ryujin, Brian S. and Scogland, Thomas RW},
  booktitle={2019 IEEE/ACM International Workshop on Performance, Portability and Productivity in HPC (P3HPC)}, 
  title={RAJA: Portable Performance for Large-Scale Scientific Applications}, 
  year={2019},
  volume={},
  number={},
  pages={71-81},
  }

@article{Guelton2015:Pythran,
	year = 2015,
	month = {mar},
	publisher = {{IOP} Publishing},
	volume = {8},
	number = {1},
	pages = {014001},
	author = {Serge Guelton and Pierrick Brunet and Mehdi Amini and Adrien Merlini and Xavier Corbillon and Alan Raynaud},
	title = {Pythran: enabling static optimization of scientific Python programs},
	journal = {Computational Science {\&} Discovery}}

@inproceedings{Okuta2017:CuPy,
  author       = "Okuta, Ryosuke and Unno, Yuya and Nishino, Daisuke and Hido, Shohei and Loomis, Crissman",
  title        = "CuPy: A NumPy-Compatible Library for NVIDIA GPU Calculations",
  booktitle    = "Proceedings of Workshop on Machine Learning Systems (LearningSys)",
  year         = "2017",
  url          = "http://learningsys.org/nips17/assets/papers/paper_16.pdf"
}

@article{Kloeckner2012:Pycuda,
author = {Kl\"{o}ckner, Andreas and Pinto, Nicolas and Lee, Yunsup and Catanzaro, Bryan and Ivanov, Paul and Fasih, Ahmed},
title = {PyCUDA and PyOpenCL: A Scripting-Based Approach to GPU Run-Time Code Generation},
year = {2012},
issue_date = {March, 2012},
publisher = {Elsevier Science Publishers B. V.},
address = {NLD},
volume = {38},
number = {3},
journal = {Parallel Comput.},
month = mar,
pages = {157–174},
numpages = {18}
}

@ARTICLE{Hunold2016:Reproducible,
  author={Hunold, Sascha and Carpen-Amarie, Alexandra},
  journal={IEEE Transactions on Parallel and Distributed Systems}, 
  title={Reproducible MPI Benchmarking is Still Not as Easy as You Think}, 
  year={2016},
  volume={27},
  number={12},
  pages={3617-3630},
  }

@article{Mann1947:On,
author = {H. B. Mann and D. R. Whitney},
title = {{On a Test of Whether one of Two Random Variables is Stochastically Larger than the Other}},
volume = {18},
journal = {The Annals of Mathematical Statistics},
number = {1},
publisher = {Institute of Mathematical Statistics},
pages = {50 -- 60},
year = {1947},
}

@INPROCEEDINGS{Karlin2013:Exploring,
  author={Karlin, Ian and Bhatele, Abhinav and Keasler, Jeff and Chamberlain, Bradford L. and Cohen, Jonathan and Devito, Zachary and Haque, Riyaz and Laney, Dan and Luke, Edward and Wang, Felix and Richards, David and Schulz, Martin and Still, Charles H.},
  booktitle={IPDPS 2013}, 
  title={Exploring Traditional and Emerging Parallel Programming Models Using a Proxy Application}, 
  year={2013},
  volume={},
  number={},
  pages={919-932},
}

@INPROCEEDINGS{Wijngaart2014:Parallel,
  author={Van der Wijngaart, Rob F. and Mattson, Timothy G.},
  booktitle={2014 IEEE High Performance Extreme Computing Conference (HPEC)}, 
  title={The Parallel Research Kernels}, 
  year={2014},
  volume={},
  number={},
  pages={1-6},
  }

@INPROCEEDINGS{Acun14:Parallel,
  author={Acun, Bilge and Gupta, Abhishek and Jain, Nikhil and Langer, Akhil and Menon, Harshitha and Mikida, Eric and Ni, Xiang and Robson, Michael and Sun, Yanhua and Totoni, Ehsan and Wesolowski, Lukasz and Kale, Laxmikant},
  booktitle={SC '14: Proceedings of the International Conference for High Performance Computing, Networking, Storage and Analysis}, 
  title={Parallel Programming with Migratable Objects: Charm++ in Practice}, 
  year={2014},
  volume={},
  number={},
  pages={647-658},
  }

\newpage




\appendices

\section{Artifact Description Appendix: Performance Evaluation of Python ParallelProgramming Models: Charm4Py and mpi4py}\label{AD}

\subsection{Abstract}
This section describes the hardware and software environment in which the results reported in this paper were obtained. Links to software used and build instructions are also given.

\subsection{Description}

\subsubsection{Check-list (artifact meta information)}

{\em Fill in whatever is applicable with some informal keywords and remove the rest}

{\small
\begin{itemize}
  \item {\bf Algorithm: Jacobi iteration, Particle-in-cell }
  \item {\bf Program: Python, Charm4Py, mpi4py, Charm++, MPI}
  \item {\bf Compilation: GCC v8.3.0; Numba v0.53.1; }
  \item {\bf Output: } Output is either a \texttt{CSV} file containing relevant output data (e.g., runtime, iteration number, process number), or data are printed to \texttt{STDOUT} and captured via pipes. Scripts to parse these data are found in the GitHub repository.
  \item {\bf Experiment workflow: } Clone and build software on the relevant platforms, run the appropriate scripts to submit jobs to the platform's management system.  
  \item {\bf Experiment customization: } Number of processes, message size, domain size, location of data, number of particles, number of warmup iterations, number of iterations, particle distribution strategy.
  \item {\bf Publicly available?: Yes. }
\end{itemize}
}

\subsubsection{How software can be obtained (if available)}\label{app:sub:software}
All software can be obtained from GitHub. All scripts and benchmark programs used to produce this paper can be found in the repository located at: \url{https://github.com/UIUC-PPL/charm4py-mpi4py-compare} in the branch \texttt{espm2\_2021}. Additional software packages can also be found on Github:
\begin{enumerate}
    \item \charmpy{} \url{https://github.com/uiuc-ppl/charm4py}
    \item \charm{} \url{https://github.com/uiuc-ppl/charm}
    \item \mpipy{} \url{https://github.com/mpi4py/mpi4py}
    \item OpenMPI \url{https://github.com/open-mpi/ompi}
\end{enumerate}

\subsubsection{Hardware dependencies} 
We describe the hardware platforms used in this paper.

\textbf{\stampede{}} is a CPU-based machine at Texas Advanced Computing Center. In our evaluation, we use the Skylake nodes. Each node is equipped with $2\times$ Intel Xeon Platinum 8160 (Skylake) @ 2.1GHz, each with 24 cores and has a memory capacity of 192GB of DDR4 RAM. Connecting the nodes is a 100Gb/s Intel Omni-Path network organized in a fat-tree topology.

\textbf{\summit{}}, located at Oak Ridge Leadership Computing Facility, is an IBM Power9 System. Each node is equipped with $2\times $ IBM POWER9 Processors, each with 44 cores. Nodes have 512GB of DDR4 memory. Additionally, each node has $6\times$ NVIDIA Tesla V100 GPUs connected with NVLink capable of 50GB/s bandwidth between GPUs. Nodes are connected with a 12.5GB/s dual-rail EDR Infiniband network in a non-blocking Fat Tree topology.

\subsubsection{Software dependencies}
In what follows, we describe the software used on both platforms, \stampede{} and \summit{}.

\textbf{\stampede{}} We compile all software with \texttt{-O3} using gcc 8.3.0. We build \charm{} on commit ID \texttt{77209f5aa} with the \mpi{} backend using the default MPI installation on \stampede{}, as we find it offers the best performance. Charm4Py commit ID \texttt{35ee630} is built against \charm{}. \mpipy{} is built on commit ID \texttt{761ac19} against the default installation of MPI; in all experiments for \mpi{} on \stampede{} we use the default installation (Intel \mpi{} v18.0.2).

\textbf{\summit{}} All software is compiled with \texttt{-O3} using gcc v8.3.1. To enable GPU-aware communication in \charm{} and \charmpy{} we build \charm{} with the UCX backend using UCX v1.11.1. UCX is itself built with gdrcopy v2.0 and libevent v2.1.12. \charm{} is built on commit ID \texttt{fa767dd9b}; \charmpy{} on commit ID \texttt{ba3e95c}

We build OpenMPI v4.1.1 with the same UCX libraries as \charm{}. \mpipy{} commit ID \texttt{23d3635} is built against this installation of OpenMPI.
Both \charm{} and \mpi{} use PMIx v3.1.1 and CUDA v10.2.89.

On both platforms, we use Python v3.8 with Numba v0.53.1. We provide the Anaconda environment file used on each platform in the GitHub repository for this paper.

\subsubsection{Datasets}
The data presented in this paper have been made available at DOI: \texttt{10.5281/zenodo.5629346}. To perform the analysis that produced the figures in this paper, the data should be extracted into a directory named \texttt{data} within the \texttt{charm4py-mpi4py-compare} repository. The analyses within the \texttt{analysis} directory can then be performed.
\subsection{Installation}
Due to space constraints, we refer the reader to the file \texttt{README.md} within the GitHub repository for the paper (Section~\ref{app:sub:software}), where we provide complete instructions for software compilation and configuration on \stampede{} and \summit{}.

\subsection{Experiment workflow}
For each benchmark ran (microbenchmarks for CPU and GPU-based messages, weak and strong scaling for CPU and GPU-based implementations of the Jacobi iteration, and the particle in cell code), scripts that submit the experiments to the scheduler are found within the \texttt{scripts} subdirectory.

\subsection{Evaluation and expected result}
Experiments may be run following software installation on the desired platforms. Scripts used to perform experiments typically require the user to change paths within the scripts to their local installation.


\subsection{Notes}
We welcome any questions and are happy to help the interested reader perform the experiments presented in this paper, or any follow-up experiments. We ask that interested parties contact Zane Fink at \textit{zanef2@illinois.edu}.

\end{document}